\documentclass[aps,a4, reprint,floatfix,prb]{revtex4-1}

\usepackage[utf8]{inputenc}
\usepackage{t1enc} 
\usepackage{graphicx}   
\usepackage{amsmath}
\usepackage{multirow}

\begin{document}

\title{Magnonic band gap and mode hybridization in continuous Permalloy film induced by vertical coupling with an array of Permalloy ellipses
}
\author{Piotr Graczyk$^{1}$} \email{graczyk@amu.edu.pl}
\author{Maciej Krawczyk$^{1}$} \email{krawczyk@amu.edu.pl}
\author{Scott Dhuey$^{2}$}
\author{Wei-Gang Yang$^{3}$}
\author{Holger Schmidt$^{3}$}
\author{Gianluca Gubbiotti$^{4}$}
\affiliation{$^{1}$Faculty of Physics, Adam Mickiewicz University, Umultowska 85, 61-614 Poznan, Poland\\
$^{2}$Molecular Foundry, Lawrence Berkeley National Laboratory, Berkeley, California 94720, USA\\
$^{3}$School of Engineering, University of California Santa Cruz, 1156 High St., Santa Cruz,
California 95064, USA\\
$^{4}$ Istituto Officina dei Materiali del CNR (CNR-IOM), Sede Secondaria di Perugia, c/o Dipartimento di Fisica e Geologia, Università di Perugia, I-06123 Perugia, Italy
}

\begin{abstract}

We investigate magnonic band structure in thin homogeneous permalloy film decorated with periodic array of elliptically shaped permalloy dots and separated by non-magnetic Pt spacer. We demonstrated experimentally formation of the magnonic band structure for Damon-Eshbach wave propagating in permalloy film  with the band gap opened at the Brillouin zone border and band splitting at smaller wavenumbers, due to the Bragg interference and interaction of propagating wave of the continuous film with a standing resonant mode of the nano-ellipses, respectively. The shape anisotropy of the permalloy nanodots allows to control the spin wave dynamics through the switch between two states of the magnetization with respect to the underneath film magnetization, thus enabling magnonic band structure reprogrammability. With numerical analysis we show, that predominant role in formation of the magnonic band structure is played by a vertical dynamic coupling between propagating wave in the film and magnetization oscillations in the nanodots.
\end{abstract}
\date{\today}
\maketitle

\section{Introduction}
Materials with designed dispersion relation are widely used to control wave propagation in electronics, microwave technology and photonics. This is realized through exploiting semiconductors or photonic crystals,\cite{Brillouin46,Elachi.1454678,Joa08,Marques2008} but also their analog for spin waves (SWs): magnonic crystals (MCs).\cite{krawczyk_2014}  Meaningful signature of periodicity is formation of the band structure and existence of the band gaps in the excitation spectra.\cite{Brillouin46,Joa08,krawczyk_2014} Although band gaps can exist for waves propagating in artificial crystals with periodicity in 1D, 2D or 3D (dimensions), to date research in magnonics is mainly focused on thin ferromagnetic films with periodic pattern in 1D or much rarely in 2D,\cite{Gubbiotti2012_a,Tacchi_2015} leaving the third dimension almost unexplored.\cite{krawczyk2013,Garst2017,Gubbiotti2018,Beginin2018}

Fabrication of thin film MCs is usually performed with patterning,\cite{krawczyk_2014,Nikitov2015} which is very receptive for defects and changes of the magnetic properties near the edges, which in turn affect the SW propagation.\cite{Klos2012,Pal_2014} Interestingly, there are also alternative ways to form SW band structure. It can be done with ion implantation, that modifies the material properties  relevant for SW propagation,\cite{Bali2014,Ruane2018,Wawro2018} and allows to create fine patterns suitable for formation of magnonic band gaps. \cite{Tahir2015,Obry2013}  Although it does not etch the film, the crystallographic structure and composition is significantly affected which usually is associated with increased damping. The other idea is based on exploitation of the vertical dynamic coupling by the stray magnetic field. In Ref.~[\onlinecite{Au2012}] the control of the SW propagation in a ferromagnetic stripe has been demonstrated by change of the magnetization orientation in the bar placed above the stripe. This mechanism was also exploited to excite SWs and to control the phase and amplitude of propagating SWs. The idea was further extended to the array of ferromagnetic nanodots deposited over the ferromagnetic film, where magnetization dynamics pumped by the microwave field in the array of dots has been used to induce/detect propagating SWs in homogeneous film.\cite{Yu2013} Only recently, the formation of the magnonic band structure in the homogeneous permalloy (Ni$_{80}$Fe$_{20}$) Py thin film by the dynamical coupling of propagating SWs with excitations confined to the  Ni stripe array has been demonstrated.\cite{Mruczkiewicz_2018} This approach is promising also for exploiting the third dimension in magnonics, which is rich with new phenomena, and promising for future SW applications.\cite{Graczyk_2018}

In this paper we investigate SW dynamics in thin Py film in presence of a 2D rectangular lattice of thin elliptical Py/Pt dots deposited on top to explore the role of the vertical coupling between the magnetization dynamics of the Py film  and the elliptical dots. We showed formation of the band structure for SWs in Py film. Moreover, the elliptical shape of the nanodots allows us to control relative magnetization orientation between the film and ellipses. We explored this property to demonstrate the re-programmability of the dipolarly induced magnonic band structure in Py film. The measurements were performed with Brillouin light scattering (BLS) technique and explained by the numerical simulations. The presented results are important step towards utilization of layered structure as basic elements for the realization of 3D magnonic systems.

The paper is organized as follows. In the next section we describe the sample structure, its fabrication, the experimental and theoretical methods used to investigate spin wave dynamics and static magnetization measurements. In Sec.~\ref{Sec:Results} we present and discuss the results of measurements and provide their interpretation based on the numerical calculations. In the last section, we summarize the results and discuss future perspectives and possible applications.

\section{Structure and methods}
\begin{figure}
\includegraphics*[width=0.45\textwidth]{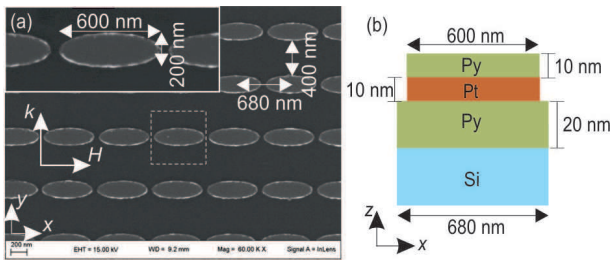}
 \caption{\label{Fig1} (a) SEM image of the  sample with indicated the main in-plane dimensions. The direction of the applied magnetic field $H$ and of the wave vector $k$, together with the Cartesian axes, are also represented by the arrows. Dashed white line marks the unit cell of the structure. (b) Schematic presentation of the unit cell in the cross-section along the $(x,z)$ plane.}
\end{figure}

Array of Py/Pt elliptical dots was fabricated on top of the continuous Py film using electron beam lithography and metal lift-off.  First, a film of 20 nm Py was evaporated using a Semicore SC600 electron beam evaporator.  Next, PMMA (poly (methyl methacrylate)) 950k C2 was spun at 5000 rpm to give a thickness of 100 nm. The pattern was then exposed with a Vistec VB300 electron beam lithography tool at 100 kV and 1 nA beam current. The sample area was 200 $\times$ 200 $\mu$m$^2$. The PMMA was then developed using a high contrast cold development process consisting of 7:3 IPA:water at 5$^o$C ultrasonicated for 100 seconds.  It was then evaporated with 10 nm of Pt (to avoid the direct contact between and the consequent exchange coupling between the Py film and dots), and 10 nm Py, and lifted off in dichloromethane. The ellipses have lateral dimension of $600\times200$ nm and are placed into chains with an edge-to-edge distance of 80 nm along the long axis of the ellipses while the inter-chain distance is of 400 nm, leading to the period $a=600$ nm and $b=680$ nm along the $y$ and $x$ axis, respectively. This corresponds to an edge of the first Brillouin Zone (BZ) along the $y$-direction of $\pi/a = \pi/600$ nm = $0.52\times 10^7$ rad/m. Scanning electron microscopy (SEM) images of the array have been obtained using a Field Emission Gun Electron Scanning Microscopy LEO 1525 (ZEISS) and are shown in Fig.~\ref{Fig1}(a). These reveal well-defined elliptical dots with uniform shape and good edge definition. A continuous (unpatterned) 20 nm Py film and array of single layer Py elliptical dots with thickness of 10 nm have been also fabricated, and used as a reference samples.

Longitudinal hysteresis loops were measured by magneto-optical Kerr effect (MOKE) magnetometry using a photoelastic modulator operating at 50 kHz and lock-in amplification.  BLS experiments from thermally excited SWs were performed in the back-scattering configuration by focusing a monochromatic laser beam of wavelength $\lambda = 532$ nm on the sample surface through a camera objective of the numerical aperture of NA = 0.24.\cite{Madami2012} The scattered light was analyzed in frequency by a (3+3)-tandem Fabry--Perot interferometer. The field dependence of SW was  measured by sweeping the applied magnetic field from +50 to -50 mT applied along the long axis of the ellipses (along the $x$ axis) at fixed wave vector $k= 1.81 \times 10^7 \text{rad/m}= 3.48 \pi/a$. The SW dispersion was then measured at fixed $H=50$ mT by sweeping the wave vector in the direction perpendicular to $H$ in the range between 0 and $2 \times 10^7$ rad/m, which allows to map the SW dispersion along the $y$ direction up to the third Brillouin zone (BZ) of the reciprocal space in the Damon-Eshbach (DE) configuration, it is when the wavevector is perpendicular to the saturation magnetization direction.

To calculate  SW spectra we solved numerically the linearized Landau-Lifshitz equation (LL) on the dynamic components of the magnetization vector $(m_y, m_z)$ in a frequency domain with damping neglected. We assume full magnetic saturation and consider the effective magnetic field $H_{\text{eff}}$ to be the sum of three terms: a dc bias magnetic field $H$ applied along the $x$ direction, an exchange field $H_{\text{ex}}$, and dynamic demagnetizing field $H_{\text{dm}}$ with components in the $(y,z)$ plane.\footnote{Due to limitation of the computation model in the calculations we neglect the static demagnetizing field coming from the magnetic charges at the elliptical nanodots edges. The possible influence of the static demagnetizing field on the SW dispersion relation is discussed in Sec.~\ref{Sec:hyb}} In calculations we assume Bloch periodic boundary conditions on the borders of the unit cell in the plane $(x,y)$ [see Fig.~\ref{Fig1}(a)]. The definition of the exchange and demagnetizing fields terms can be found in Ref.~[\onlinecite{Mruczkiewicz2013}]. We added also a small magnetic anisotropy term of value 10 mT parallel to the $H$ in nanodots to fit confined modes with the experimental data. We solved the LL equations in 3D space using finite element method for the in-plane unit cell and with large free space above and below the Py along the $z$-axis.\footnote{We used  COMSOL 4.3a software. The maximum element size was: along the \textit{z} direction: 5 nm within elipse and layer, 15 nm in the surrounding; along \textit{x} and \textit{y} directions: 40 nm within elipse and layer, 60 nm in the surrounding.} From solutions of the LL equation we found frequencies $f$ of the SWs for successive wavenumbers $k$ along the $y$-axis, i.e., the magnonic band structure. We also obtained the spatial distribution of the SW amplitude $(m_y, m_z)$ and from the $m_z(x,y,z)$ we calculated the BLS intensity of the SWs:\cite{Jorzick1999}
\begin{equation}
I_{\text{BLS}} \propto \left|\int_{\text{volume}} m_z(x,y,z) e^{iky}dx dy dz\right|^2. \label{Eq:I_BLS}
\end{equation}
To take into account small penetration depth of light into the metal we excluded from integration,  the part of the Py film just under elliptical nanodots. For further details concerning the computation method we refer to Ref.~[\onlinecite{Mruczkiewicz2013}].

\begin{figure}
\includegraphics*[width=0.3\textwidth]{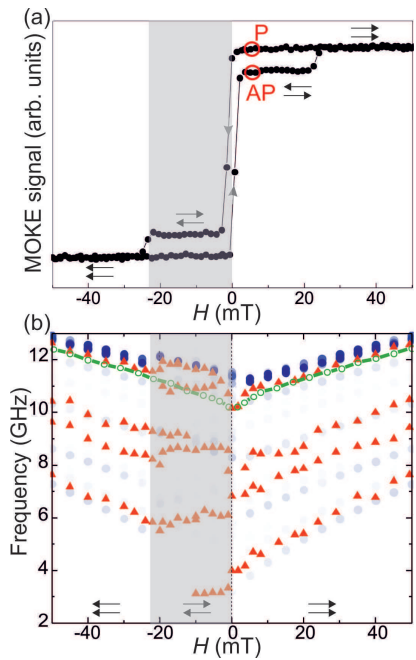}
 \caption{\label{Fig:2} (a) MOKE loop measured with the magnetic field  along the long axis of the ellipses (the $x$ direction). Red circles  mark 5 mT field at which the SW dispersion in P and AP states were measured and are shown in Fig.~\ref{Fig:7}(a) and (b), respectively. (b) The SW frequency at $k= 3.48 \pi/a$  in dependence on $H$ oriented in the $x$ direction. The BLS measurements were started from saturation in the high positive values of $H$ and decreasing the field toward negative saturation following the descending branch of the MOKE loop.  SW frequencies, derived from the measured BLS spectra, are marked by red triangles. The results of numerical calculations are shown with the blue colored dots, with the color intensity proportional to the calculated BLS intensity. The green line with the empty dots shows the BLS measurements on the reference 20 nm thick continuous Py film.  }
\end{figure}

\section{Results}\label{Sec:Results}

\subsection{Field dependence}\label{Sec:field}
The measured hysteresis loop is shown in Fig.~\ref{Fig:2}(a). It was measured with the magnetic field directed along the long axis of the ellipses starting from large positive fields. In saturation the magnetization in the film and nanodots is parallel (P state). At small negative external magnetic field values the magnetization in the Py film  switches to the opposite direction but the magnetization in the elliptical nanodots remains along the same direction thus providing anti-parallel orientation of the magnetization (AP state). The magnetization remains at the AP orientation for $H$ covered the plateau visible in Fig.~\ref{Fig:2}(a), i.e., between -2 and -22 mT. With higher magnetic field the magnetization inside the ellipses also reverses reaching saturation at negative fields below -25 mT. The stabilization of two different magnetization configurations in the film will be exploited in the further part of the paper.

 To study the evolution of mode frequency in dependence of the magnetic field we measured a sequence of BLS spectra in the DE configuration for a fixed $k = 3.48 \pi/a$  with decreasing the magnetic field magnitude starting from the saturation state at 50 mT. The results are shown in Fig.~\ref{Fig:2}(b) with the red triangles. The simulation results are superimposed with the color map where the intensity of the color is proportional to the calculated BLS intensity according with Eq.~(\ref{Eq:I_BLS}).
 At $H =50$ mT we observe 4 modes and their frequencies decrease with decreasing the magnetic field in good agreement with the simulation results. The most intensive line at 12.5 GHz is the DE mode of the Py film.  Softening of the DE mode near the transition to AP state (near $H = 0$) is  visible in BLS measurements but it is not reconstructed in simulation. This is because of the assumption of the full saturation required by the model. Near the magnetization reorientation of the film  we observe sudden change of the frequencies which nicely confirms drop in the magnetization curve in the MOKE measurements. In the range of field values where there AP state is stable, we distinguish the two main types of  $f(H)$ dependencies. In the first type $f$ increases with increasing $\left| H \right|$ along the negative values while in the second type $f$ decreases or weakly depends on $H$. We can relate the bands with increasing $f$ to the waves traveling in Py film. This observation is compared with the numerical results in Fig.~\ref{Fig:2}(b). The analysis of the mode profiles shows that the most intensive mode (12 GHz at -20 mT)is the DE type of wave and it has the frequency increasing with $\left| H \right|$ starting from -2 to -20 mT.    
In the AP state the orientation of the magnetization in nanodots is opposite to the magnetic field direction, thus we attribute the branches of decreasing frequency (e.g., the band at 6 GHz) to the  oscillations confined to the nanodots. The frequencies just follow  decrease of the internal magnetic field in nanodots in this case. Nevertheless, this dependence is affected by the coupling with the Py film, the different strength of the interaction results in various slopes in the $f(H)$ dependencies. At fields of magnitude larger than $\left| 22 \right |$ mT, the magnetizations are in the P state again, and all four modes increase their frequency with increasing magnetic field magnitude towards negative values. 

Interestingly, in the P state the SW frequency of the DE type of wave is very close to the frequency of the DE wave in the reference Py film (see green line in Fig.~\ref{Fig:2}(b)). However, the difference becomes significant (up to 1 GHz) in the AP state at low field values. This points at possibility for valuable control of the DE mode frequency by the change of the magnetization orientation between the film and nanodots---the property investigated in details in further part of the paper. 

\subsection{Magnonic band structure and hybridization}\label{Sec:hyb}

\begin{figure}[h!]
\includegraphics*[width=0.3\textwidth]{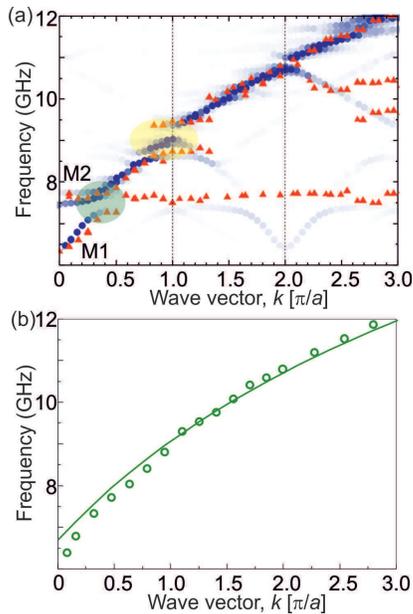}
 \caption{\label{Fig:3} (a) Dispersion relation of SWs in DE configuration measured with BLS (red triangles) and obtained from simulations (color map) at the magnetic field $H = 50$ mT, i.e., in the P state. The intensity of the colors in the numerical data relates to the calculated BLS intensity of the SWs. The vertical dashed lines mark the BZ boundary and center. Green and yellow areas indicate the hybridization and tha Bragg band gap regions, respectively. (b) Dispersion relation of SWs measured in the reference sample, 20 nm thick Py film. The green line is to guide experimental points.}
\end{figure}

 The measured dispersion relation of SWs at saturation ($H = 50$ mT) for waves propagating perpendicular to the field is shown in Fig.~\ref{Fig:3}(a) with red triangles. We distinguish very intensive line (M1) with the group velocity decreasing with increasing $k$. This band starts around 6.34 GHz and reaches 12 GHz at $k = 3 \pi / a$, which is very close to the DE wave in the reference sample shown in Fig. \ref{Fig:3}(b). Near the first BZ boundary we found the splitting of the DE band into two excitations 8.73 GHz and 9.46 GHz [the area marked in yellow in Fig. \ref{Fig:3}(a)], which we attribute to the opening of the magnonic band gap of the DE mode due to Bragg interference condition. We also noticed another DE band splitting at lower frequencies and smaller wavenumbers, between $7.27$ and 7.73 GHz at  $k = 0.4 \pi / a$ (at the area marked in green).  The frequency of this gap is close to the almost non-dispersive line (M2) visible through the whole range of wavevectors, its frequency oscillates around 7.6 GHz. The dispersion relation obtained from the simulations (color map in Fig.~\ref{Fig:3}(a)) matches well with the BLS data, and also demonstrates the two magnonic gaps in the spectra. 

\begin{figure}[h!]
\includegraphics*[width=0.46\textwidth]{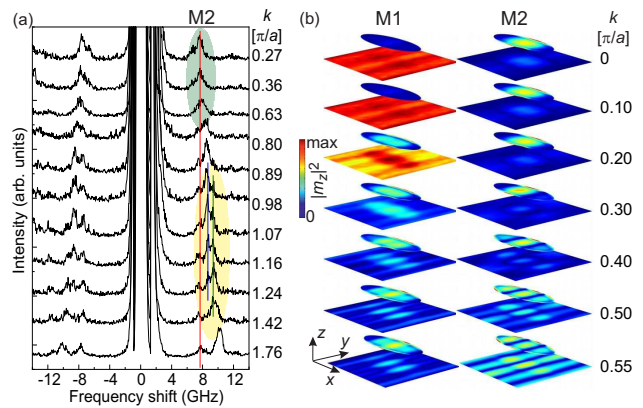}
 \caption{\label{Fig:4} (a) The BLS spectra for selected wavevectors at $H= 50$ mT from which the dispersion in Fig.~{Fig:3}(a) is formed. The yellow and green areas in (a) mark the $k$ range of the band gap opening at the BZ border and due to hybridization of the DE wave with the confined modes in elliptical nanodots, respectively [see also Fig.~\ref{Fig:3}(a)]. (b) The amplitude of SWs $m_z^2$ in arbitrary units across the mid-planes of the elliptical nanodots and the plain film for the two modes M1 and M2 at various $k$ values to demonstrate the hybridization of SWs near the band gap opening due to hybridization.}
\end{figure}

To interpret the low frequency gap  we show in Fig.~\ref{Fig:4}(a) a sequence of BLS spectra measured at different $k$ values for an applied magnetic field of +50 mT. We do not observe any frequency asymmetry between the Stokes and anti-Stokes side of the spectra as could be expected in presence of Dzyaloshinskii-Moriya  interaction near the Pt/Py interface.\cite{Tacchi2017} This is because of the rather large thickness of the Py layers. 
 At $k< 0.8\pi/a$, all the measured spectra present a peak doublet where the frequency position of one peak significantly depends on $k$ following the frequency evolution of the DE mode in the reference Py film, while for another one, marked by the vertical red line (M2 mode), the frequency is almost constant. Interestingly, for $k<0.5 \pi/a$ the mode at constant frequency has largest intensity which decreases with increasing $k$, while the intensity of dispersive mode significantly increases with $k$. This change in the relative intensity of the peak together with the fact that their frequency difference decreases until the two peaks almost merge one with the other, and then increases again, are the characteristic signature of a hybridization and repulsion, i.e.,  two modes at  $k$ value close to  fulfill phase matching condition change their character from propagative to standing nature and vice versa. 

These experimental findings can be understood by inspection of the $k$-vector evolution of the calculated spatial profiles of the modes in the two Py layers shown in Fig.~\ref{Fig:4}(b). 
In the low $k$-vector range the lowest frequency mode (M1) mainly exists with an extended profile in the continuous Py film and with a negligible spin precession amplitude within the ellipses. This mode M1 resembles the Kittel mode of the Py film with uniform spatial profile. On the contrary, in the same $k$-vector range, the mode M2 represents the quasi-fundamental SW excitation of the Py ellipses without nodal lines and the amplitude maximum in the ellipse center. Close to the crossing point of the M1 and M2 dispersion (where $k$ ranges from 0.3 to 0.5 $\pi/a$) the spatial distribution in the two layers is rather similar, i.e., the profiles in the Py ellipses is essentially replicated within the underneath Py film with mode M1 which loses its extended character. Eventually, for $k>0.5 \pi/a$ the mode M1 has a largest amplitude in the ellipses while mode M2 is mostly confined in the continuous film with an extended character and several amplitude oscillations, thus supporting the idea of modes which exchange their character.

\begin{figure}[h!]
\includegraphics*[width=0.3\textwidth]{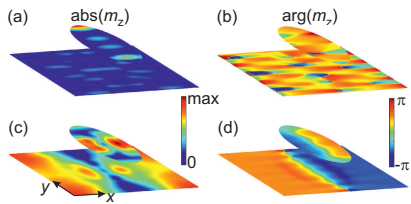}
 \caption{\label{Fig:5} The calculated amplitude (a and c) and the phase (b and d) of SW at the magnonic band gap at the BZ boundary: at $f = 8.73$ GHz (a and b) and $f = 9.46$ GHz (c and d)}.
\end{figure}

For $k>0.9 \pi/a$ [see, Fig.~\ref{Fig:4}(a)] an additional peak appears in the spectra at higher frequency with respect to the doublet discussed above. Just around the edge of the BZ the frequency of the two peaks at higher frequency remain constant. This is a signature, that opening of a magnonic band gap occurs at the edge of the first BZ. Eventually, on further increasing $k$ one mode disappears while the frequency of on one start to increase with $k$. 

Calculated profiles of SWs (their amplitude and phase) from the magnonic band gap edges at the BZ boundary are shown in Fig.~\ref{Fig:5}. The amplitude distribution has rather complex oscillating pattern due to hybridization with other excitations present at similar frequencies. Nevertheless, it is clear that the SW at the BZ boundary, from the bottom of the Bragg gap, has only one nodal line in the unit cell (it is along the direction of wave propagation). The line is located in the film as well as in the middle of the nanodots. Interestingly, the magnetization oscillations in nanodot and film are shifted by $180^\circ$ for mode below the band gap [Fig.~\ref{Fig:5}(d)], but they oscillate in-phase for the mode just above the magnonic band gap [Fig.~\ref{Fig:5}(d)]. This reminds the band gap between acoustical and optical vibrations in the bi-component atomic chain, with different atoms oscillating in-phase and out-of-phase in respective bands.\cite{Kittel} However, in our structure the in-phase oscillations are for the mode of higher frequency. It is, because the dipolar dynamical interaction between film and dots is minimized by the out-of-phase oscillations (determined by their out-of-plane components of the magnetization) in both subsystems. 

\begin{figure}[h!]
\includegraphics*[width=0.4\textwidth]{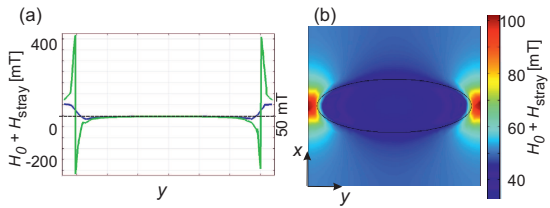}
 \caption{\label{Fig:6} The static magnetic fields distribution in the cross-sections of the unit cell: (a) along the middle axis (the $y$ axis) of the nanodot (green line) and in Py film below this line (blue line), (b) in the $(x,y)$ plane in the middle of the film. The field is a superposition of the external magnetic field and the stray magnetostatic field, $H_0 + H_{\text{stray}}$, $H_0 = 50$ mT.}
\end{figure}

Let's discuss now a possible influence of the inhomogeneous static stray magnetostatic field in the Py film induced by elliptical dots on the magnonic band gap. This field was neglected in computations of the band structure, but we calculated it from the static computations by solving Gauss equation. This stray field (with the external field of 50 mT) is shown in Fig.~\ref{Fig:6}. We see, that although the field is inhomogeneous, its amplitude in the Py film is rather small reaching maximal values close to 100 mT only at small areas along the $y$ axis which corresponds to the separation between the elliptical dots. Also the average values of the stray field are small (7 mT in ellipses and close to 0 in Py film). These features point at rather minor influence of the static dipolar field on the SW properties of the Py film  (DE mode). To prove this, we performed additional calculations of the magnonic band structure in thin Py film with inhomogeneous (periodic) external magnetic field, equal to the static stray field created by ellipses [shown in Fig.~\ref{Fig:6}(b)]. The inhomogeneity of the field introduces the folding back effect and formation of other eigensolutions of small BLS intensity. However, it results in the band gap at the BZ boundary of only 50 MHz which is small compared to 730 MHz wide gap due to dynamic coupling. Thus, we conclude, that the static field does not affect significantly the band structure of DE SWs and the dominating role is played by the vertical (between the two Py layers, through Pt) dynamical dipolar coupling.

\subsection{Re-programmability}

\begin{figure}[h!]
\includegraphics*[width=0.3\textwidth]{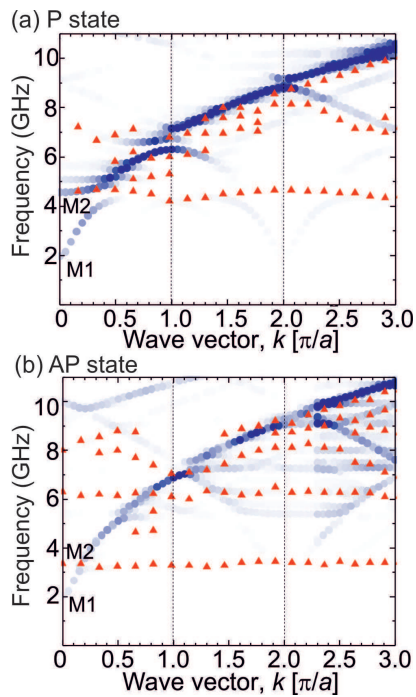}
 \caption{\label{Fig:7} Dispersion relation of SW propagating perpendicular to the magnetic field of value 5 mT aligned along the $x$-axis for (a) parallel (P) and (b) anti-parallel (AP) state of the magnetization in nanodots and the film. The red triangles mark the  measured BLS data, the color map show the numerical results. The color intensity of the numerical points indicates the calculated BLS intensity. Horizontal dashed lines mark the BZ boundary and center.}
\end{figure}

The discussion in Sec.~\ref{Sec:field} allows us to propose the re-programmable band structure in 2D MCs in which by the change of the magnetic field history we can reach P or AP states of the magnetization orientation. We select $H =5$ mT for the demonstration and  prepared the system in these two states, see red circles in Fig.~\ref{Fig:2}(a).  The measured and calculated magnonic band structures for P and AP states are shown in Fig.~\ref{Fig:7}  (a) and (b), respectively. We found the spectra for P and AP state to be significantly different.

\begin{figure}[h!]
\includegraphics[width=0.4\textwidth]{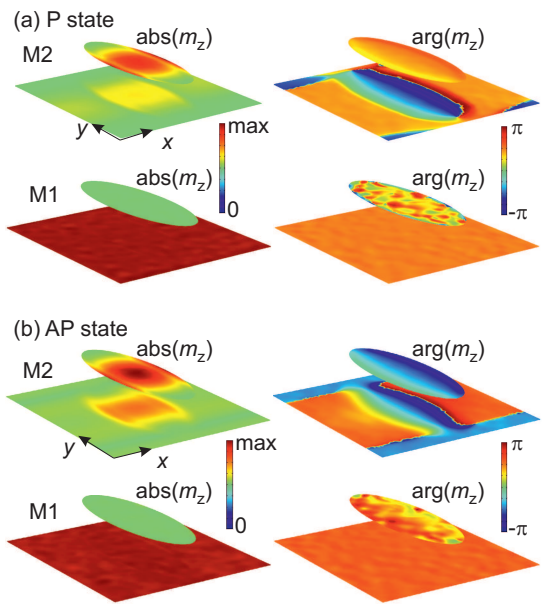}
 \caption{\label{Fig:8} The calculated amplitude $|m_z|$ and the phase $arg(m_z)$ for SWs in the P (a) and AP (b) state of the magnetizations. The calculations are for the DE wave (M1) and fundamental excitation in nanodot (M2) marked in Fig.~\ref{Fig:7} at $k = 0$. }
\end{figure}

In the  P state [Fig.~\ref{Fig:7}(a)] the spectra is similar to Fig. \ref{Fig:3}, but shifted to the lower frequencies. At $k = 0$ the DE band is not experimentally detected, in simulation it starts at 1.98 GHz. Measured is the first standing mode in nanodots which extend over all range of $k$ with frequency varied from 4.2 to 4.7 GHz. At higher $k$ values the DE mode dominates in the spectra. The profile of the DE mode and of the first standing mode in the nanodot are shown in Fig.~\ref{Fig:8}(a) for the P state. As in the previous results [Fig.~\ref{Fig:4}], the DE mode (M1) concentrates amplitude mainly in the film while the standing mode of the nanodot (M2) forces oscillations also in the Py film. In the film the amplitude is weaker than in the nanodot, but oscillations in film have phase shifted by $\pi$ with respect to the oscillation in nanodot.

 In the AP state, Fig.~\ref{Fig:7}(b) the SW spectrum is richer, than in the P state. Through the whole $k$ range the two nondispersive lines are visible in BLS measurements (the lowest standing mode is measured at around 3.4 GHz in agreement with calculations). The DE mode becomes visible only  from the second BZ. The profile of the DE (M1) and the first standing mode of the nanodot (M2) in the AP state at $k=0$ are shown in Fig.~\ref{Fig:8}(b). The oscillations in the DE mode are very similar to that one in P state, however the M2  preserves significant differences. There, the amplitude enhanced in Py film and the $z$ component of the magnetization are with the same phase in nanodot and in the film, while $m_y$ components are  in anti-phase. This points out at the opposite direction of precession in both sub-systems, which is due to the opposite orientation of the magnetization. 

These results are very important, since they show that the magnonic band structure can be significantly shifted in frequency by choosing the appropriate ground state, while the applied magnetic field is fixed at a specific value.

\section{Conclusions}
In summary, we have investigated experimentally and theoretically SW propagation in homogeneous thin Py film decorated with the rectangular array of thin Py nanodots of the elliptical shape placed on the top of the film, and separated by a Pt spacer to avoid the exchange coupling. We have demonstrated the formation of the magnonic band structure and opening of the magnonic band gaps. Based on numerical analysis we concluded, that the main mechanism of the coupling is based on the dynamic coupling of the confined oscillations in the nanodots with the propagating waves in the film. This type of interaction allows for formation of the Bragg magnonic band gap for the waves propagating in the film plane and perpendicularly to the direction of the magnetization. Moreover, the strong dynamic interlayer dipolar coupling results also in anti-crossing the dispersion of the propagating wave with the standing excitation when the phase matching conditions are fulfilled. The coupling can be modified by change the Pt thickness. Making Pt layer separating the Py film and nanodots below the spin diffusion length, at ferromagnetic resonance frequency of the nanodot additional coupling via spin pumping effect can be exploited. 
These properties, together with the re-programmable magnonic band structure, are  important contribution to 
magnonics and spintronics. 
  
\vspace{-0.6cm}
\section*{acknowledgments}
\vspace{-0.5cm}
This work was supported by the National Science Foundation grants ECCS-1509020 and DMR-1506104,
National Science Centre of Poland grant UMO-2012/07/E/ST3/00538 and from the EU Horizon 2020 Research and Innovation Programme GA No. 644348(MagIC). Work at Molecular Foundry was supported by the Office of Science, Office of Basic Energy Sciences, of the U.S. Department of Energy under Contract No. DE-AC02-05CH11231. Scanning electron micrograph have been taken at the LUNA laboratory, Department of Physics and Geology of the University of Perugia. 

\bibliography{bibliography}

\end{document}